\documentclass[aps,prapplied,12pt,superscriptaddress,notitlepage]{revtex4-2} 
\usepackage[utf8]{inputenc}
\usepackage{amsmath, amssymb, bm}      
\usepackage{graphicx}                   
\usepackage{wrapfig}                    
\usepackage{array}                       
\usepackage{booktabs}                   
\usepackage{setspace}                   
\usepackage{placeins}                   
\usepackage{chngcntr}                   
\UseRawInputEncoding
\usepackage{xr}                         

\usepackage{tikz}                       
\usetikzlibrary{trees}                  
\begin{document}
\title{Loss-driven gain enhancements driven by topological singularities in non-Hermitian photonic crystals defects}

\author{Daniel Cui}\affiliation{Department of Materials Science and Engineering, University of California, Los Angeles, Los Angeles, CA 90095, USA}
\author{Aaswath P. Raman}
\email{aaswath@ucla.edu}
\affiliation{Department of Materials Science and Engineering, University of California, Los Angeles, Los Angeles, CA 90095, USA}
\affiliation{California NanoSystems Institute, University of California, Los Angeles, Los Angeles, CA 90095, USA}

\begin{abstract}

We show that purely lossy defects in one- and two-dimensional non-Hermitian photonic crystals can induce transmission matrix singularities not accessible with lossless defects. These singularities in turn can enable dramatic enhancement in overall system gain not accessible through conventional means. We further show that the underlying mechanism behind the loss-induced gain enhancement is due to the resonances being located specifically at topological branch cut singularities in the reflection coefficient with nontrivial winding numbers. The resulting resonances can exhibit exceptionally high quality factors in excess of $\sim 10^4$. Our work highlights the counterintuitive role of loss in engineering singularities in the gain response in non-Hermitian systems and its connection to topological phenomena in photonic systems. 

\end{abstract}
\maketitle 

\section{Introduction}
Photonic structures have been widely use to enhance nonlinear optical response, improving gain and enabling high quality factors. In conventional photonic crystals, coherent effects enabled by tuning material parameters such as thickness and permittivity can lead to characteristic resonances and stop bands that are widely exploited for devices, including distributed feedback lasers \protect\cite{coldren2012DFB}. Lossless defects in photonic crystals have additionally been shown to localize particular modes either within the photonic crystal bandgap \cite{joannopoulos2008molding} or within the continuum of states as a BIC \cite{vaidya2021defectBIC}, with resonant enhancement being exploited for a range of applications. 

Recent advances in non-Hermitian photonic systems have further demonstrated that judicious arrangements of gain and loss can access new behaviors of relevance to many optical device applications. Parity-time symmetric systems \cite{ozdemir2019parity,bender1998real} and the observation of exceptional points \cite{zhou2019exceptional,ali2009exceptional_singularity} in the photonic bandstructure have been observed due to non-Hermitian Hamiltonians being able to host real spectra and the coalescence of their associated eigenstates \cite{bender1998real}. Such behavior is unique to such systems and not present in their Hermitian counterparts. New applications have also leveraged these novel phenomena and unique combinations of gain and loss systems in PT-symmetric lasers \cite{hossein2014ptlaser,seker_single-mode_2023} and in photonic integrated circuits \cite{pereira2024NHPIC}. Enhancement of emission in nonreciprocal systems \cite{zhong_nontrivial_2021,reisenbauer_non-hermitian_2024} and Anderson localization in disordered photonic systems \cite{piao2022anderson,Longhi2024anderson} have also benefited from non-Hermitian architectures. While the role of defects has been investigated in non-Hermitian photonic systems, including non-Hermitian photonic crystals, the role of lossy defects in such systems has not been explored. 

In this context, we note that new frameworks have also been introduced to describe the underlying physical mechanisms behind the nontrivial phenomena observed in non-Hermitian photonic systems. One such framework involves looking at the formation of exceptional point (EP) degeneracies \cite{miri_exceptional_2019} which are unique to non-Hermitian systems, particularly as the matrix describing the response becomes defective and non-diagonalizable due to coalescence of both the eigenvalues and eigenstates \cite{heiss_physics_2012}. In contrast, eigenvalue degeneracies in Hermitian systems do not necessarily yield EP's as the eigenstates are not guaranteed to coalesce as well. These specific kind of degeneracies are called Diabolical Points (DP) \cite{berry_diabolical_1984, RevModPhys.68.985}. EP's and their higher dimensional generalizations \cite{cerjan2016exceptionalcontour,wang2023nhlattices} have thus yielded novel ways of analyzing and engineering the spectral responses in non-Hermitian systems for example in parity-time symmetric systems \cite{ruter2010observation,chong_PT_laser_absorber_2011}, coherent perfect absorption phenomena \cite{liu2023cpa,chong2010cpalaser,ergoktas2024topthermalinterface,liu2023cpa} and loss-induced lasing \cite{peng_loss_lase_2014,Liertzer_Pump_EP_2012}. 

Another framework used in the study of non-Hermitian systems is the examination of the poles and zeros of the system's response functions. Such functions, including the scattering matrix $S(\omega)$ and the transmission matrix $T(\omega)$, physically describe how the electromagnetic fields and energy are transferred from the input to the output channels \cite{binkowski2024NHPoles&Zeros}. Mathematically, it is known that these response functions can be expanded into a product of zeros and poles in the complex frequency plane using the Weierstrass theorem  \cite{vanKampen1953SingOrigin}. Studying the spectral response in this way has been used to explain phenomena such as Wood's anomalies \cite{Maystre2012WoodsAnomaly,hessel1965woods} and more recently for wavefront engineering in gradient metasurfaces displaying $2\pi$ phase accumulation \cite{colom2023branchcut, lawrence_high_2020}. Such singularities have previously been examined in S-matrix \cite{guo_singular_top_2023,Wang_CPA_EP_2021} and the Hamiltonian defined by temporal coupled mode theory \cite{peng_loss_lase_2014,Valero_EP_BIC_TCMT_2025}.

In this article, we show that introducing lossy defects in non-Hermitian gain photonic crystals result in exceptional points degeneracies purely from examining the eigenvalue spectrum in the transmission matrix.  We then show that these exceptional points are positioned within the vicinity of reflection poles that have nontrivial topology which can be further optimized to enhance the gain response. In particular, instead of acting like undesirable absorption, we demonstrate that a lossy defect placed in a non-Hermitian photonic crystal can induce multiple orders of magnitude enhancement in gain compared to the pure periodic system without loss. We characterize the nature of this phenomenon using the integer winding number of the complex reflection phase, finding that peak gain values occur precisely at the point of a topological phase transition in the system at $2\pi$ branch cut phase singularities. We further generalize this behavior to a 2D non-Hermitian photonic crystal system, demonstrating that the loss-enhanced resonances in the 1D and 2D systems are both fundamentally symmetry-protected quasi-BIC's and attain very high quality factors at the topological phase transition point. 

\section{Results}

\subsection{Exceptional Point Enhancement}

We begin by considering the gain response of a model one-dimensional periodic non-Hermitian photonic crystal system that prior work suggests can support an exceptional point. In this system we consider 50 unit cells, each of length $a$ and consisting of a lossless layer of length $0.8a$ and permittivity $\varepsilon_a$ = 2, and a gain layer of length $0.2a$ with permittivity $\varepsilon_b$ = 2 - 0.1i (Fig. 1(a) top). When examining the reflection and transmission spectrum of this periodic system, as calculated by the Transfer Matrix Method (TMM), we find a sharp resonant peak in the gain spectrum, reaching a maximum value of $\sim 10^4$ (Fig. 1(b)). Examining the transfer matrix $T$ for a two port system and imposing reciprocity conditions results in the condition that $\det T = 1$. This in turn allows us to directly solve for the eigenvalues of the transfer matrix
\begin{equation}
    \lambda = \frac{1}{2} [(t_{11} + t_{22}) \pm \sqrt{(t_{11}+t_{22})^2 - 4 \det T}
\end{equation}
Given the potential role of an exceptional point in enabling this gain peak we consider the constraint imposed by the exceptional point on the eigenvalues to find the following expression
\begin{align}
    \begin{split}
(t_{11} + t_{22}) + \sqrt{(t_{11}+t_{22})^2 - 4 detT} &= (t_{11} + t_{22}) - \sqrt{(t_{11}+t_{22})^2 - 4 detT} \\
&=\sqrt{[Tr(T)]^2 - 4} = 0
    \end{split}
\end{align}

\begin{figure}[ht]
    \centering

    \includegraphics[width = 6.5in]{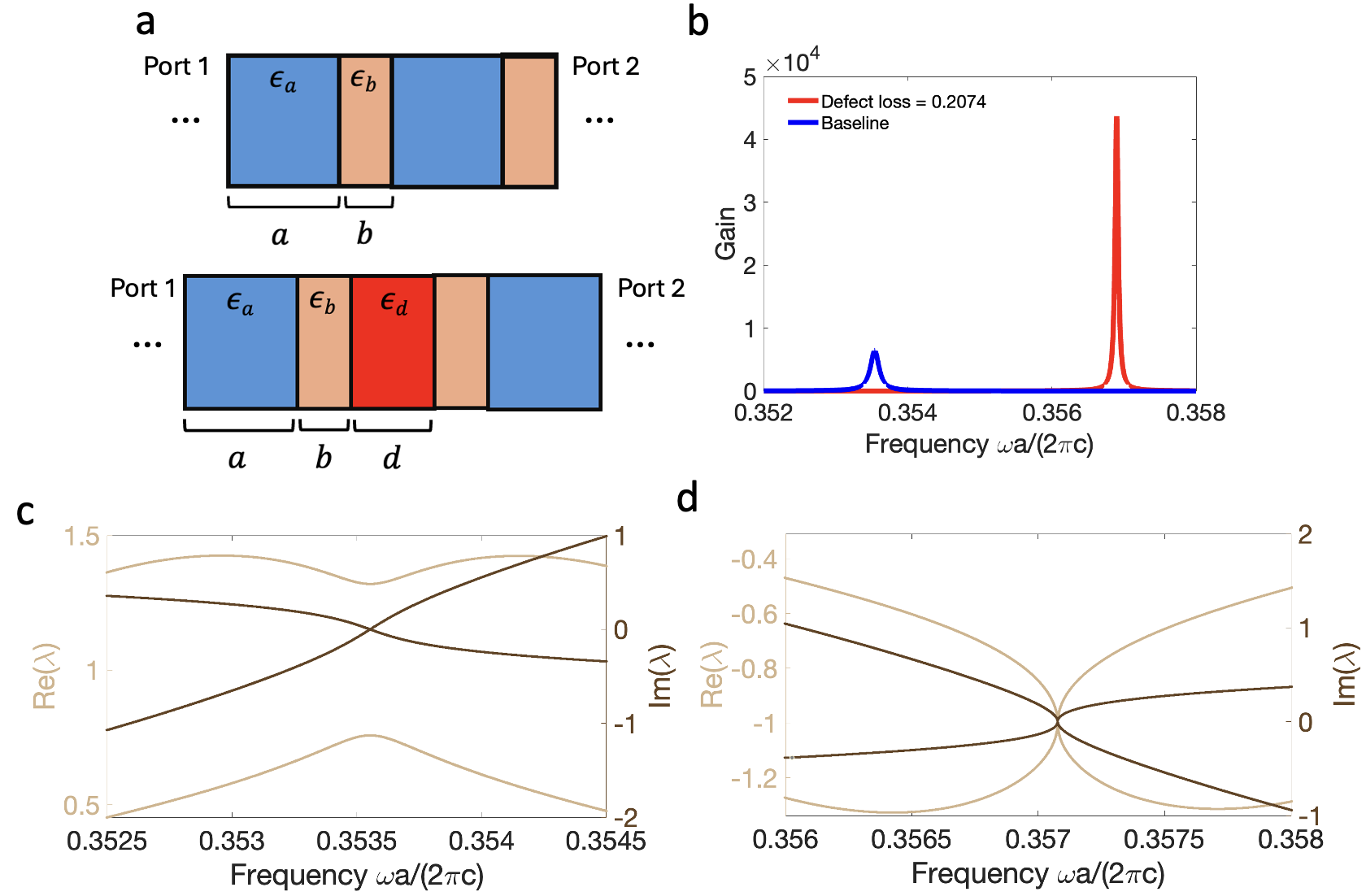}
    \caption{(a) Schematics of the baseline periodic and defected 1D non-Hermitian photonic crystals on the top and bottom respectively. The baseline periodic system has 50 unit cells of length d of lossless $\varepsilon_a$ = 2 with length a = 0.8*d and gain layers of $\varepsilon_b$ = 2-0.1i with length b = 0.2*d. The defected structure is inversion symmetric about the center of the lossy defect layer of $\varepsilon_d = 2 + \varepsilon_d''*i$ with length d. 25 unit cells of the original baseline periodic system are located on the left of the defect. Those 25 unit cells on the left are then reflected across the center of the defect to create the right half of the structure. (b) Gain as measured by $ |1-R-T| $ for the baseline periodic structure, defected system at a loss value of 0.2074.(c), (d) Plots of the real and imaginary parts of eigenvalues of the transmission matrix for the baseline periodic and lossy defect structures respectively.}

    \label{fig:1D Result}
\end{figure}

The eigenvalues $\lambda_1 $ and $ \lambda_2$ of $T$ at each frequency of the structure shows the absence of any exceptional point within the vicinity of the gain peak frequency. Further analysis of the norm squared of T matrix elements $t_{11}$ and $t_{22}$ reveal that $|t_{11}|$ is approximately zero (Fig. S1(a)). Since both $R$ and $T$ are defined to be proportional to $\frac{1}{|t_{11}|^2}$, the two attain large values which is consistent with the observation of the gain peak. However, $|t_{22}|$ is slightly above 2 (Fig. S1(b)). Based on the exceptional point condition derived in Eq. (2), $|t_{22}|$ must be equal to 2 to be at an EP. Physically, this implies a non-unitary $T(\omega)$ near the gain peak which is already given by the non-Hermiticity of the structure through the gain layers. To arrive at an EP singularity, one must then require that the overall amplification of the backwards going wave cannot be arbitrarily high. Instead it must be balanced to reach this condition. 

We thus examine the role of a centrally placed defect on enhancing the gain response of the system by inducing a singularity in the transmission matrix. In particular, we consider a lossy point defect of thickness $d$ and complex permittivity $\varepsilon_d = 2 + \varepsilon_d''*i$ where $\varepsilon_d''$ denotes the magnitude of the imaginary permittivity (Fig. 1(a) bottom). The structure with the defect is constructed to be inversion symmetric about the center of the defect with 25 unit cells on either side of the lossy defect. As shown in Fig. 1(b), gain peaks can be found within $\frac{\omega a}{2\pi c}$ of 0.34 to 0.37 for both the baseline system and the defected structures for a defect thickness of $d = 0.5a$ where $a$ is the unit cell length of the baseline periodic system. Furthermore, the gain peak itself is significantly larger. This enhancement can be understood by noting that, as the loss $\varepsilon_d''$ is tuned, the real parts of the T-matrix eigenvalues coincide at the same frequency as the imaginary parts. Examination of $|t_{11}|$ and $|t_{22}|$ show that indeed the EP condition is satisfied with $|t_{11}|$ remaining approximately 0 while $|t_{22}|$ approaches 2. 

\begin{figure}[ht]
    \centering
    \includegraphics[width = 6in]{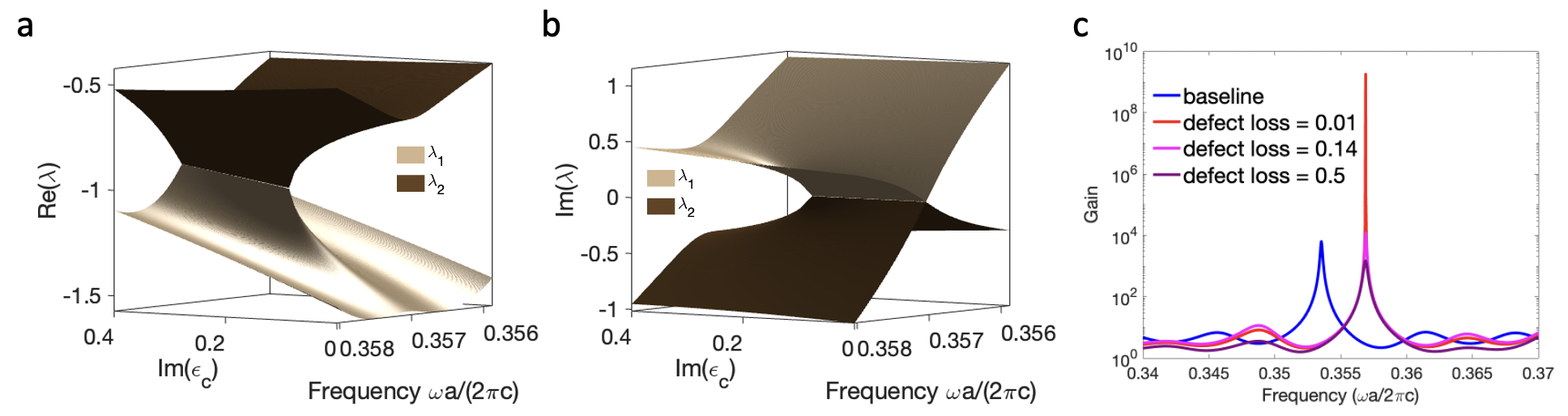}
    \caption{(a), (b) Surface plots of  the real and imaginary parts of the eigenvealues of the 2-by-2 transmission matrix respectively in the defect loss and $Re(\omega)$ parameter space. An EP forms from a degeneracy in the real and imaginary parts coinciding at around a loss of 0.2074 and frequency of 0.3571. (c) Gain for the baseline periodic structure, defected system at a loss value of 0.01, defected system at the critical defect loss value of 0.14, and defected system at a defect loss value of 0.5.}
    \label{fig:1D_Phase Singularity}
\end{figure}

To elucidate the full eigenvalue spectrum in the vicinity of the gain peak frequency, we plot the real and imaginary parts of $\lambda_1$ and $\lambda_2$ in the defect loss ($Im(\varepsilon_c))$ and frequency parameter space (Figs. 2(a) and 2(b)). We note that, for this particular structure, the EP degeneracy forms at a defect loss of 0.2074 which is the onset of $Re(\lambda_1) = Re(\lambda_2)$. However, beyond this loss value, the imaginary parts of the eigenvalues split. Remarkably, at the loss value that generates the EP, the gain peak of the lossy defect structure is $4$ x higher than the baseline periodic structure without a lossy defect (Fig. 1(b)). One might think that being at the EP in this system has the particular physical consequence of enhancing the gain response. However, increasing the strength of the defect loss layer to a smaller value to 0.14 leads to even more dramatic gain enhancement by several orders of magnitude to $> \sim 10^8$ (Fig. 2(c)). Beyond this value, the gain decreases, suggesting that an optimal value of the loss in the defect is optimizing the singularity-like behavior of gain peak. This singularity must thus correspond to the poles of reflection coefficient and similarly in transmission. 

\begin{figure}[ht]
    \centering
    \includegraphics[width = 6in]{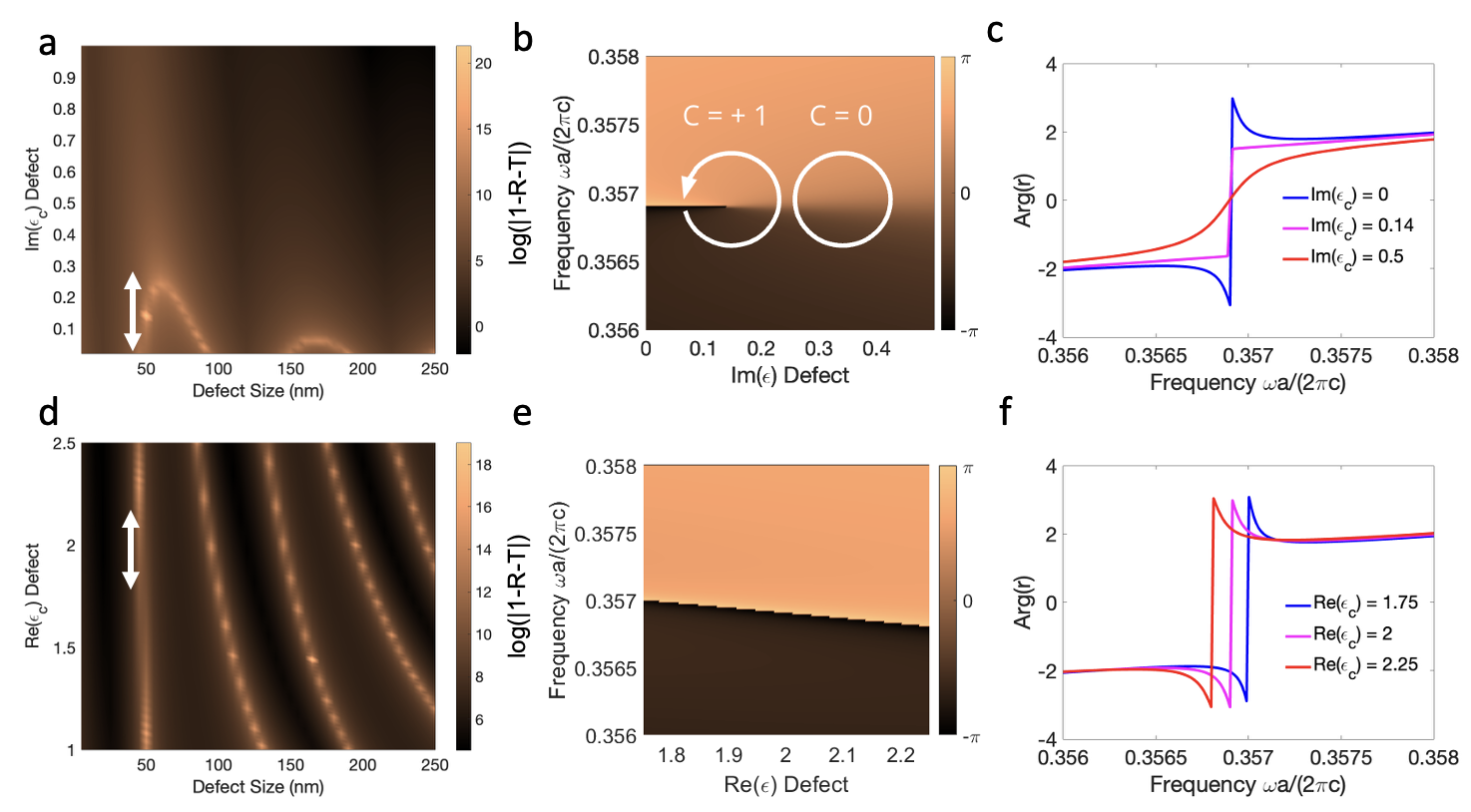}
    \caption{(a), (b), (c) Maximum gain achieved in the defect structure by sweeping defect loss vs defect size, complex reflection phase plotted over $Re(\omega)*a/(2\pi c)$ where a is the unit cell length and c is the speed of light vs defect loss, and cross sections of the reflection phase color plot at various defect loss values. A branch cut singularity appears starting at $\varepsilon_c'' = 0$ and goes until $\varepsilon_c'' = 0.14$ after which the complex reflection phase is no longer singular. The branch cut singularity also has a topological charge of C = +1. (d) Maximum gain achieved in the defect structure by sweeping $ Re(\varepsilon) $ of the defect with no loss vs defect size. Complex reflection phase plotted over $Re(\omega) a/(2\pi c)$ vs $Re(\varepsilon_c)$ of the defect without any defect loss with color plot and cutlines in (e) and (f) respectively. The complex reflection phase exhibits a discontinuity for all values of $Re(\varepsilon_c)$. Maximum gain is achieved when the loss in the defect is around 0.14 at a reflection pole.}
    \label{fig:2D_result}
\end{figure}

\subsection{Topological Phase Singularity Enhancement}

To further characterize this behavior, we calculate the gain for a range of values of both the defect loss as well as defect layer thickness (Fig. 3(a)). Bands of high gain values show an unusual character, with a specific optimal loss value existing for a given defect layer thickness. This is in notable contrast to the gain behavior observed with lossless defects when the real part of the defect's permittivity and thickness are varied (Fig. 3(d)). The resulting ranges of values for permittivity and layer thickness are consistent with a typical coherent interference effect. However, we note that using a lossy defect not only produces a larger gain enhancement than a lossless approach, but the fundamental character of maximal gain values follows a distinct pattern. These peak values also attain high Q factors of ~ $4\times 10^4$ due to being quasi-BICs (Figs. S2(a), S3(a), and S3(b)).

In order to elucidate the physical phenomenon behind the unique capabilities of lossy defects to induce gain enhancement, we turn to examining the phase of the complex reflection coefficient as it varies with frequency. In particular, we fix the defect thickness at 50 nm and examine a range of $\varepsilon'_d$ and $\varepsilon''_d$ values (Fig. 3(a) and 3(b)). The reflection phase is taken to be $Arg(r)$ where $r$ is the complex reflection coefficient in the TE polarization. This then allows us to calculate a winding number for this system, following analogous approaches to characterizing topological properties of bound states in the continuum \cite{wu2024BICphase}, exceptional points \cite{colom2023branchcut}, coherent perfect absorption, the Su-Schrieffer-Heeger model of the topological phases of polyacetylene \cite{asboth_short_2016,Henriques2020Tamm,su1979solitons}, and topological defects in the XY model of spins \cite{mermin1979topdefect,gingras1996defects}. We use the following formulation of the integer winding number:

\begin{equation}
C = \frac{1}{2\pi} \oint d\phi,  C \in Z
\end{equation}
where $\phi$ is the complex reflection coefficient as defined above and the convention for the path taken for line integration is taken in the direction of increasing reflection phase. In most regions of frequency space, the winding number assumes a trivial value due to there being a completely continuous definition of $Arg(r)$. In the presence of singularities however, the winding number can take on nontrivial integer values. 

By modulating the loss $\varepsilon_d''$ in the lossy defect of the 1D system while keeping the real part of $\varepsilon_d$ constant, we observe a nontrivial winding number (Fig. 3(b)). As shown in both the complex reflection phase color plot and the slices of the color plot at various magnitudes of loss in Figs. 3(b) and 3(c), the complex reflection phase is discontinuous and singular up to a loss value of 0.14. Beyond 0.14, the complex reflection phase becomes continuous, indicating the presence of a branch cut singularity. Taking an anticlockwise $2\pi$ phase loop around the branch cut, the winding number takes on a value of +1 which is a nontrivial value of the topological charge $C$ of the system. Furthermore, it is important to note that the the point at which the phase singularity becomes continuous at a loss value of 0.14 is also precisely the point where the gain is maximized in Fig. 2(c). At loss values beyond this critical point, the gain decreases. Thus, the gain resonance is topologically protected and remains so until a topological phase transition is made from the singular to non-singular regions of the frequency vs $Im(\varepsilon_d)$ parameter space as indicated by the winding number changing from a nontrivial $C = +1$ to trivial $C = 0$ transition (Fig. 2(a)). The nontrivial effect that defect loss has on the topology of the complex reflection phase can also be clarified by modeling the 1D point-defect system as a Fabry-Perot cavity (Figs. S4(a) and S4(b)). Specifically, the loss that leads to the transition in the complex reflection coefficient corresponds to a reflection pole at the same frequency.

In contrast, modulating $Re(\varepsilon_d)$ in the lossless defect case leads to very different behavior. As seen in the color plot in Fig. 3(e), and the slices at various defect loss values (Fig. 3(f)), the phase transition is completely discontinuous, indicating the presence of reflection poles as well. However, no continuous integration path exists for suitably defining the integer winding number in equation 2 and, consequently, no topological phase transition can be found in this case. The role of $Re(\varepsilon_d)$ in the defect is in stark contrast to the role of $Im(\varepsilon_d)$. In the latter, a topological phase transition as defined by nontrivial value of $C$ is induced by the defect loss whereas the modulation of the former lacks such a transition despite both scenarios exhibiting reflection poles.  

\begin{figure}[ht]
    \centering
    \includegraphics[width = 6 in]{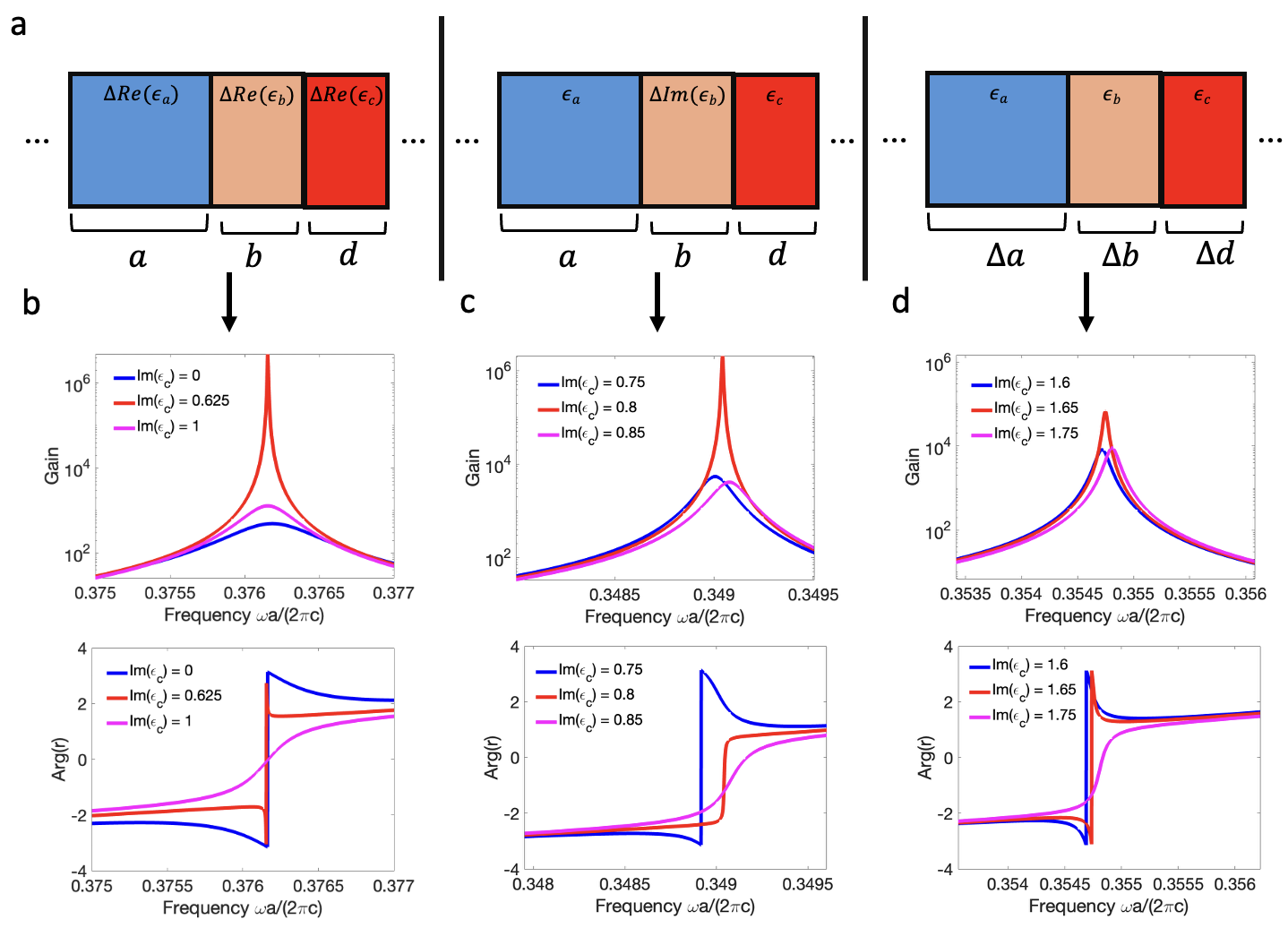}
    \caption{(a) Examination of the role of various structural parameters on the gain response and reflection phase singularity in 3 ways: changing the real part of the permittivities in all the layers, changing the gain in the gain layers, and adjusting the thicknesses in all layers.  Gain spectra and complex reflection phase cutlines corresponding to modification in (b), (c), and (d) respectively. The red curves show the maximum gain achieved by positioning the gain peak at a reflection pole through increasing the defect loss.  }
    \label{fig:2D_gain}
\end{figure}

The unique role of defect loss in gain enhancement is not restricted to the particular parameters of the 1D model system that we have discussed so far. For example, changing the real parts of the permittivities of the layers from 2 to 1.8 leads to a shift in the resonance frequency position (Fig. 4(a) left and 4(b) top). However, additional loss is needed in the lossy defect layer to obtain nontrivial gain enhancement and the transition in the winding number of the complex reflection phase (Fig. 4(b) top and bottom). A similar effect is seen when doubling the gain in the gain layer from 0.1 to 0.2 (Fig. 4(a) middle). Additional loss in the defect is required to realize the same amount of gain and induce the winding number transition (Fig. 4(c) top and bottom). Lastly, we show the effect of changing the layer thicknesses (a = 70 nm, b  = 30 nm, and d = 75 nm) in Fig. 4(a) right. By increasing the fill factor of the gain layers in the periodic layers surrounding the defect, a larger loss value is needed in the defect layer to optimally enhance the gain peak and trigger the winding number transition (Fig. 4(d) top and bottom). 

\subsection{2D Loss-Driven Gain Enhancement}

\begin{figure}[ht]
    \centering
    \includegraphics[width = 5.5in]{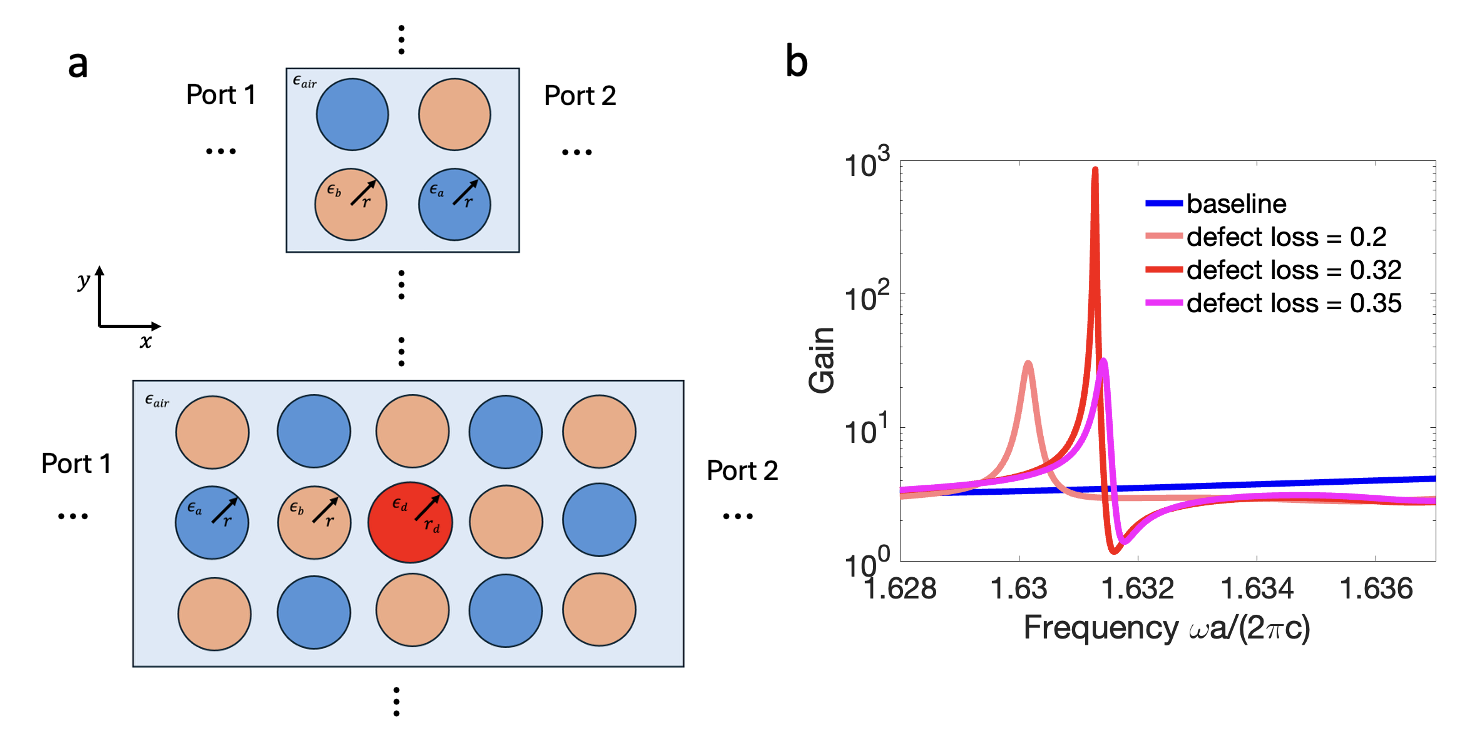}
    \caption{(a) schematics of the baseline periodic and point-defected 2D non-Hermitian system. The baseline periodic system is shown on top with 12 unit cells of length a in the x-direction while infinitely periodic in the y-direction using Floquet periodic boundary conditions. Each unit cell is composed of air surrounding lossless rods of $\varepsilon_a = 2$ and gain rods of $\varepsilon_b = 2-0.1i$ with radii $0.17a$. The defected case is shown below which contains a lossy point defect composed of a rod with $\varepsilon_d$ and radius $r_d = 0.3*a$. The defect is surrounded by 6 unit cells of the baseline periodic system on the left and right extents and made to be $C_4$ symmetric about the defect. (b) Gain vs frequency plots of the baseline periodic crystal compared against the lossy line defected case at loss values of 0.2, 0.32, and 0.35.}
    \label{fig:2D_phase}
\end{figure}

 The results observed thus far in 1D systems can also be extended to two-dimensional non-Hermitian system. We consider a baseline periodic 2D non-Hermitian photonic crystal with 12 unit cells that repeat in the $x$ direction of rods surrounded by air similar to the 1D baseline periodic structure (Fig. 5(a)). Within each unit cell, there is one lossless rod of $\varepsilon_a = 2$ and a rod with gain of $\varepsilon = 2 - 0.1i$. The rods have identical radii $0.17a$ where $a$ is the unit cell length. While the photonic crystal is finite in the x-direction, Floquet periodic boundary conditions are imposed to make the system infinitely periodic in the $y$-direction. In the defected case, a geometry similar to the 1D point-defected system is used where 6 unit cells surround either side of a lossy or lossless defect rod defined by a permittivity $\varepsilon_d = \varepsilon'_d + \varepsilon''_di$ and radius $r_d = 0.3a$, resulting in $C_4$ rotational symmetry about the center of the point defect.

\begin{figure}[ht]
    \centering
    \includegraphics[width = 5.5in]{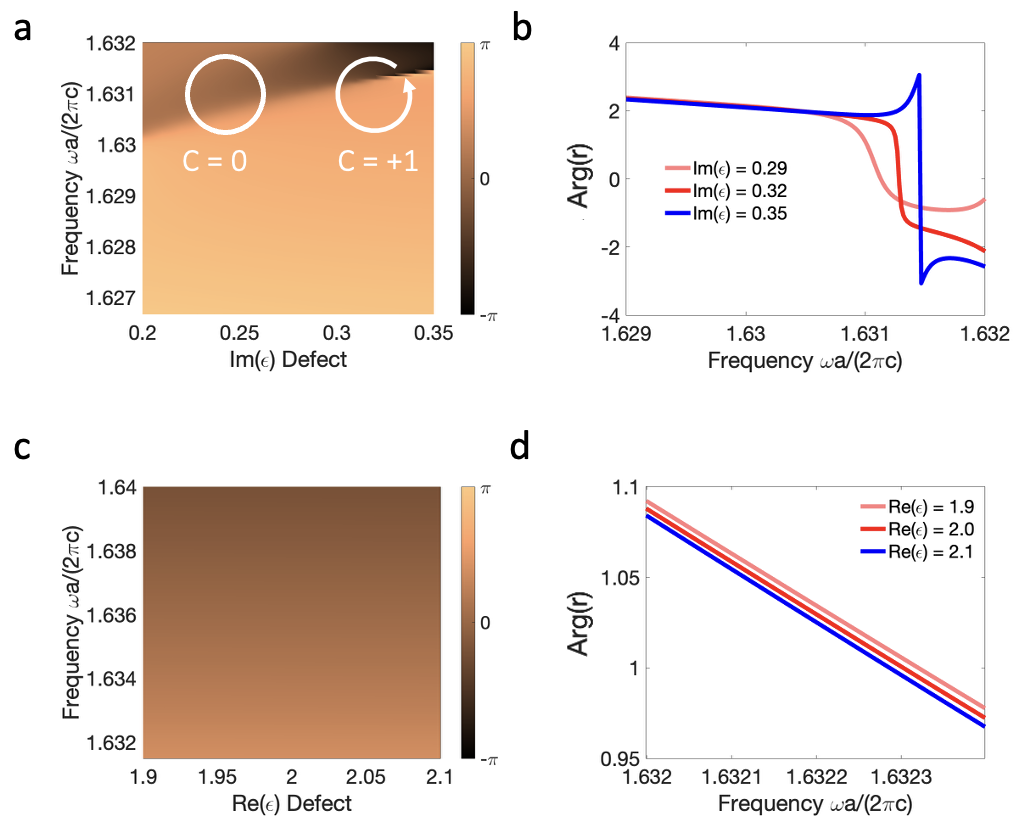}
    \caption{(a), (b) Complex reflection phase as quantified by $Arg(r)$ vs $Im(\varepsilon_d)$ and (c), (d) $Re(\varepsilon_d)$ at 0 loss of the point defect respectively. Increasing $Im(\varepsilon_d)$ leads to a transition from a winding number of 0 to +1, while the complex reflection phase remains continuous with changing $Re(\varepsilon_d)$ in the absence of loss.}
    \label{fig:quasi_BIC}
\end{figure}

As in the 1D system, increasing the loss in the point defect up to a critical value leads to multiple orders of magnitude gain enhancement (Fig. 5(b)) with extremely high quality factors due to the gain peaks corresponding to quasi-BICs (Figs. S3(b), S4(c), and S4(d)). This behavior can be understood by examining the integer winding of the complex reflection phase in the frequency vs $\varepsilon_d$ parameter space. Modulating the loss within the defect induces a topological phase transition around a critical loss value of $\varepsilon''_d = 0.32$ (while maintaining $\varepsilon_d' = 2$) as a transition from a nontrivial winding number of $C$ = 0 to a trivial winding number of $C$ = +1 (Fig. 6(a)). This is due to the presence of a branch cut phase singularity where the transition point from nontrivial to trivial winding is where the gain enhancement is at a maximum (Fig. 6(b)). By contrast, changing only the real part $\varepsilon'_d$ does not induce any phase transition as the complex reflection phase is completely continuous throughout the frequency range of interest (Fig. 6(c) and 4(d)). Thus, we observe again that the loss-induced gain enhancement from the line defect is due to the topological protection of the gain resonance as inferred from the integer winding number as the topological invariant. 

In summary, we have demonstrated that purely lossy defects can play the unexpected role of inducing both exceptional point and reflection pole singularities exhibiting nontrivial topology in non-Hermitian photonic crystals. In the case of reflection pole singularities, gain enhancement can be achieved and optimized through a topological phase transition in 1D and 2D non-Hermitian photonic crystals. Our work highlights the unique role that material loss and defects can play in non-Hermitian photonic structures. More broadly, it highlights further opportunities for engineering singularities to maximize performance in nonlinear photonic systems.  

\section{Acknowledgments}
This material is based upon work supported by the National Science Foundation under grants No. ECCS-2146577. We thank Dr. Alexander Cerjan for helpful discussions related to this work. 

\bibliography{Ref}

\counterwithin{figure}{section}

\clearpage

\end{document}